\documentclass[aps,prl,showpacs,amsmath,twocolumn,superscriptaddress]{revtex4}
\usepackage{graphicx}
\usepackage{color}
\usepackage{amsmath}
\usepackage[american]{babel}

\begin{document}

\def\K{{\bf{K}}}
\def\Q{{\bf{Q}}}
\def\Gbar{\bar{G}}
\def\tk{\tilde{\bf{k}}}
\def\k{{\bf{k}}}
\def\Kp{{\bf{K}}^{\prime}}
\def\tp{t^{\prime}}
\def\Jeff{J_{\textrm{\textit{eff}}}}

\title{Insensitivity of superconductivity to disorder in the cuprates}

\author{A. Kemper}
\email[]{kemper@qtp.ufl.edu}
\affiliation{University of Florida, Gainesville, FL 32611, USA}
\author{D. G. S. P. Doluweera}
\affiliation{University of Cincinnati, Cincinnati, Ohio, 45221, USA}
\author{T.A. Maier}
\affiliation{Oak Ridge National Laboratory, Oak Ridge, Tennessee, 37831, USA}
\author{M. Jarrell}
\affiliation{University of Cincinnati, Cincinnati, Ohio, 45221, USA}
\author{P.J. Hirschfeld}
\affiliation{University of Florida, Gainesville, FL 32611, USA}
\author{H-P. Cheng}
\affiliation{University of Florida, Gainesville, FL 32611, USA}

\date{\today}

\begin{abstract}
Using a dynamical cluster quantum Monte Carlo approximation, we
investigate the effect of local disorder on the stability of
$d$-wave superconductivity including the effect of electronic
correlations in both particle-particle and particle-hole channels.
With increasing impurity potential, we find an initial rise of the
critical temperature due to an enhancement of antiferromagnetic
spin correlations, followed by a decrease of $T_c$  due to
scattering from impurity-induced moments and ordinary
pairbreaking.  We discuss the weak initial dependence  of $T_c$ on
impurity concentration found in comparison to experiments on
cuprates.
\end{abstract}
 \maketitle

{\em{Introduction.}}
Disorder is an essential feature of the
superconducting cuprates. Crystal growth procedures lead
generically to  defects such as grain boundaries, atomic site
switching,  and vacancies. Additional disorder, often in the form
of oxygen or other charged defects, is almost always introduced
away from the CuO$_2$ plane upon doping the parent compound from
the Mott insulating state.
%
This last type of disorder may be
responsible for local nanoscale electronic inhomogeneity  in the
superconducting state of the cuprate Bi-2212 indicated by scanning
tunnelling spectroscopy (STS)
experiments\cite{t_cren_00,c_howald_01,s_pan_01,k_lang_02}.
These experiments show modulations of the local gap near the impurity sites
on the order of the correlation length\cite{t_nunner_2005,t_nunner_2006}.
A recent experiment imaging high-energy resonances thought to be
the dopant atoms themselves  shows a strong positive correlation
of the magnitude of the local spectral gap with the locations of
the dopants, leading to suggestions that the origin of the
observed gap modulations are caused by atomic scale variations in
the pairing interaction\cite{t_nunner_2005,t_nunner_2006}.

When impurities like Zn and Ni are substituted for Cu in the
CuO$_2$ plane, or planar defects created by electron irradiation,
 superconductivity is suppressed\cite{g_xiao_90,t_chien_91,s_tolpygo_96a,f_rullier-albenque_00}.
Because the screened Coulomb potential due to these defects is
very short range\cite{lwang_05}, such impurities are frequently
modeled by pointlike ($\delta$-function) scatterers.  The
expected form for the suppression of superconductivity in the BCS
theory of $d$-wave superconductors  is then identical to the
expression given by Abrikosov and Gor'kov\cite{a_abrikosov_1961}
for magnetic impurities in $s$-wave superconductors (see, e.g.,
Ref.\onlinecite{a_balatsky_06}). However, experimentally
 a significantly slower initial slope of the $T_c$ suppression is
 observed. For example,  Tolpygo et al\cite{s_tolpygo_96a} reported a
suppression  2-3 smaller than the AG curve\cite{s_tolpygo_96a} in
resistivity measurements of YBCO films.  Other unusual deviations
from AG behavior have been observed at larger disorder levels; for
example, an electron irradiation study\cite{f_rullier-albenque_03}
on optimally doped YBCO reported a linear  behavior in $T_c$ vs.
resistivity over the entire $T_c$ range.

Theoretically, several possible effects beyond Abrikosov-Gor'kov
(AG) theory have been explored.  A numerical mean field study
of disordered $d$-wave superconductors\cite{m_franz_97} including the
self-consistent suppression of the order parameter around each
impurity site showed deviations from the AG result.  Several
authors attempted to account for the slowness of the $T_c$
suppression by assuming that the scattering potential of planar
impurities was extended, or
anisotropic\cite{g_haran_96,g_haran_98,m_kulic_97,m_kulic_99}.
Recently, Graser et al.\cite{s_graser_07} calculated both $T_c$ and
the impurity resistivity $\rho$ within a consistent model of
extended potential scatterers, and concluded that the unusual
$T_c$ vs. $\rho$ behavior seen in cuprate experiments should be
attributed to strong correlations or strong coupling corrections
to BCS theory.  In general, the effect of correlations on the
structure and scattering of quasiparticle states in a disordered
$d$-wave superconductors is still an open and very important
question for cuprates and other unconventional superconductors.

One interesting consequence of disorder in a correlated electron
host is impurity-induced magnetism: nuclear magnetic resonance
measurements indicate the formation of magnetic moments upon
chemical substitution of a nonmagnetic impurity for a
Cu\cite{h_alloul_91,a_mahajan_94}. This was corroborated by
calculations of the magnetic spin susceptibility, which displays
Curie-Weiss behavior upon impurity doping (see e.g. Ref.
\onlinecite{t_maier_02b}). Several aspects of  theory and
experiment in connection with  disorder-induced magnetism in
cuprates and 1D spin systems have recently been
reviewed in Ref. \onlinecite{h_alloul_08}. 
While most of the theoretical work on these questions has been
confined to the normal state, the quasiparticles deep in the
$d$-wave superconducting state are also affected.  Mean-field
calculations utilizing the Gutzwiller
approximation\cite{a_garg_2006} suggest that the effects of
disorder on the density of states are suppressed in the presence
of strong correlations, specifically near the nodes and at low
energies. Similar effects in the density of states are also
recovered in calculations where correlations are treated in a
simple Hartree-Fock scheme by Andersen et al.\cite{b_andersen_07},
who found however that although the effects of disorder on the
density of states were indeed weakened,  some unusual effects
outside the framework of BCS theory were also present, e.g. the
breakdown of universal transport in d-wave
superconductors\cite{p_lee_93}.

In this paper, we aim to understand some of the effects of
disorder on the suppression of the transition to $d$-wave
superconductivity. First, a small concentration of weak impurities
is shown to cause an increase in the effective antiferromagnetic
exchange coupling, which enhances superconductivity within the
Hubbard model. At the same time, the disorder causes pairbreaking,
which tends to suppress $T_c$. As the impurity potential is
increased, the pairbreaking overcomes the enhancement of $J$
causing a decrease in $T_c$, which continues until it saturates
when the unitary limit is achieved. We suggest that these effects
may partially account for the observed slow suppression of $T_c$
by disorder in the cuprates.

{\em{Formalism.}}
The Hamiltonian of our model is
\begin{equation}
\label{eq:Hamiltonian}
H= -t\sum_{\langle ij\rangle\sigma}\ c^\dagger_{i\sigma} c_{j\sigma}
  + U \sum_i n_{i\sigma} n_{i -\sigma}+  \sum_i V_i n_{i\sigma}
\end{equation}
where $ c^\dagger_{i\sigma} (c_{i\sigma})$ creates (destroys) an
electron with spin $\sigma$ at site $i$, and $
n_{i\sigma}=c^\dagger_{i\sigma} c_{i\sigma}$. Here ${\langle ij
\rangle}$ denotes  nearest neighbor sites $i$ and $j$, $U$ denotes
the on-site Coulomb repulsion and and $t$ is the nearest-neighbor
hopping amplitude. The impurity is modeled as a potential $V_i = V$
on a single site. We shall give a brief review of the method (see Ref.
\onlinecite{t_maier_02b} for further details).

To study 
(\ref{eq:Hamiltonian}), we employ the dynamical cluster
approximation (DCA)\cite{m_hettler_98,m_hettler_00,th_maier_05a}.
The DCA is a  dynamical mean-field theory which self-consistently
calculates the self-energy on a cluster of size $N_c$ embedded in
a host.  Correlations on the cluster are treated explicitly.
Interactions beyond the cluster scale are dealt with on a
mean-field level within the self-consistent host. With increasing
cluster size, the DCA systematically interpolates between the
single-site dynamical mean field (DMFT)\cite{a_georges_96} and the
exact result. Cluster dynamical mean field calculations (including
the DCA) of the Hubbard model are found to correctly obtain many
of the features of the cuprates, including a Mott gap and strong
AF correlations, d-wave superconductivity and pseudogap
behavior\cite{th_maier_05a}. To solve the cluster problem, we use
a quantum Monte Carlo (QMC) algorithm\cite{m_jarrell_01c}, and
employ the maximum entropy method\cite{m_jarrell_96a} to calculate
the real frequency dynamic spin susceptibility. The sign problem in 
QMC is small for the values of $U$ considered, and is therefore not an issue for
the calculations presented here.

The result of the QMC calculation depends on the disorder
configuration, but $T_c$ is determined by the average Green's
function, which we compute in the following way. For a
concentration $x$, contributions with $m$ impurities on the
cluster are weighted by a combinatoric factor $x^m(1 - x)^{N_c -
m}$.  It is reasonable, for small concentrations $(x < 1/N_c)$, to
consider only those configurations with zero or one impurities.
For the zero and one impurity case, the combinatoric factors
expand to $1 - x N_c$ and $x N_c$, respectively.  We can then
write the disorder average
\begin{equation}
\label{eq:disorder_average}
G^c_{ij} = x N_c G^c_{1,ij} + (1 - x N_c) G^c_{0,ij}
\end{equation}
 where $G^c_{m,ij}$ is the cluster real space
Green's function for $m$ impurities. The disorder-averaged Green's
function is then used to continue the DCA algorithm.

 To determine the critical temperature $T_c$, we extrapolate the
pair-field susceptibility $\chi_d(T)$\cite{th_maier_05a}, and note
that the system enters the superconducting state when $\chi_d(T)$
diverges. To interpret the results we present below, we will also
need to calculate the induced magnetic moment $m$.  This is done
using   a method introduced by Krishna-murthy et
al\cite{h_krishnamurthy_80a}. We note that the square magnetic
moment in the low-temperature limit is proportional to $T$ times
the magnetic susceptibility. To study the effect of the impurity,
we subtract the pure susceptibility, and arrive at
\begin{equation}
\label{eq:magnetization}
m^2_{induced} \propto T (\chi^c_1 - \chi^c_0)
\end{equation}
where $\chi^c_1$ and $\chi^c_0$ are the susceptibilities of a
cluster with a single impurity and a homogeneous cluster,
respectively.

{\em{Results.}}
We carry out DCA/QMC calculations using the $N_c=16$, type
A\cite{d_betts_99} cluster for the Hamiltonian in
Eq. \ref{eq:Hamiltonian}, fix the doping at ~10\% and let $U=4t$. 
Estimates for $T_c$ have been shown to be robust against cluster size effects\cite{th_maier_05b}.
Furthermore, we have investigated a possible finite size effect by observing the change zero frequency
spin-spin correlation function (not shown), which was found not to deviate appreciably from the clean cluster
beyond the first nearest neighbour \textemdash indicating that the finite size effect does not
play a significant role on the quantities we report.\\
We first investigate the $d$-wave superconducting transition
temperature $T_c$ and the induced moment of the system as a
function of impurity potential $V$ for various values of the impurity concentration.

\begin{figure}[ht]
\includegraphics[width=3.2in, clip = true]{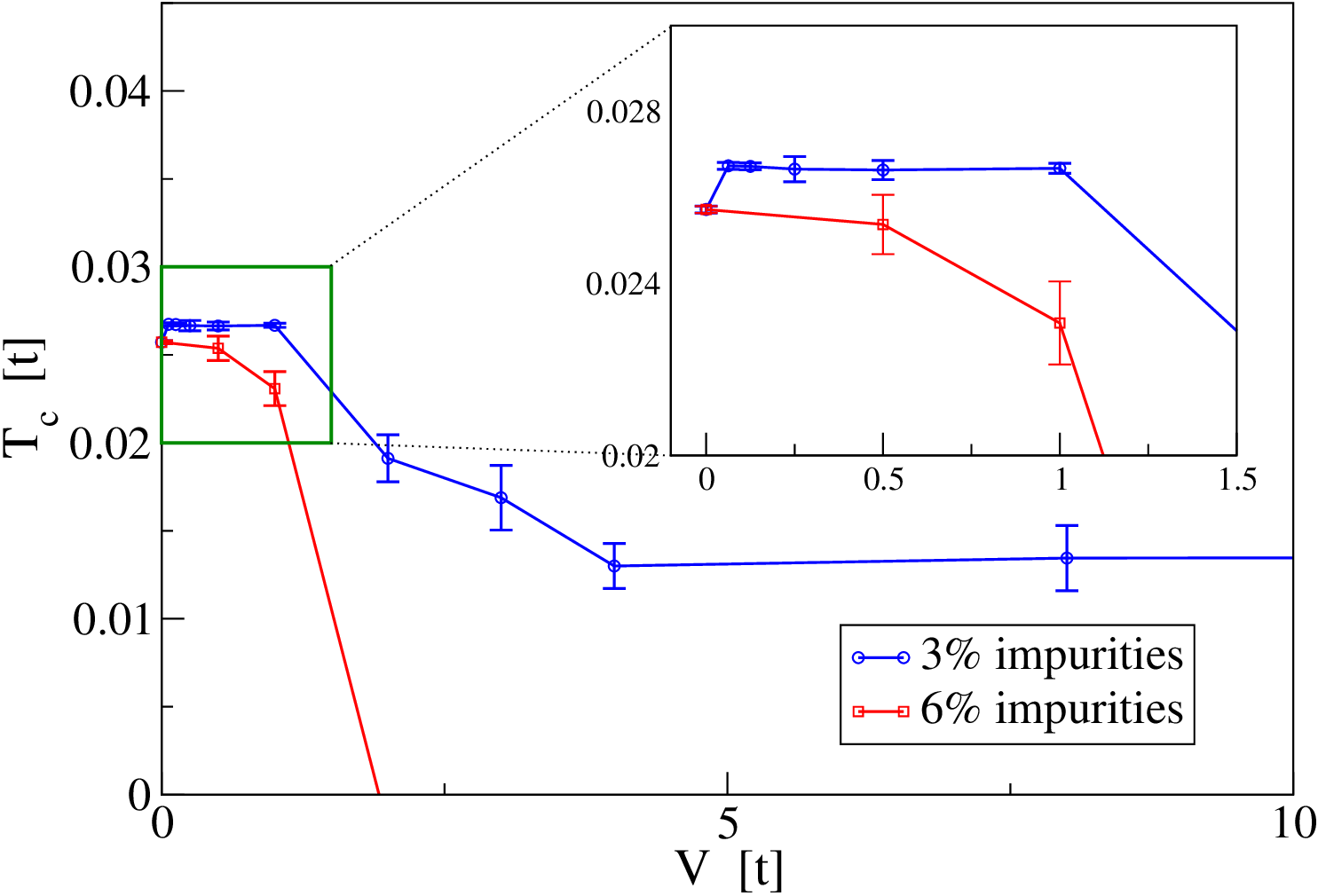}
\caption{The critical temperature $T_c$ as a function of impurity potential for $N_c$ = 16 and
$U$ = $4t$, at impurity concentrations $x = 3\%$ and $x = 6\%$. Error bars
are calculated from the extrapolation of the pair-field susceptibility\cite{t_maier_02b}.
Inset: Blowup of the region of small impurity potential.}
\label{fig:Tcv}
\end{figure}
Our first significant finding is the  initial weak increase of
$T_c$ in the region $0 < V \leq t $ for 3\% impurity concentration
(Fig. \ref{fig:Tcv}).  This is completely unexpected from the
point of view of AG theory, where any concentration of impurities
of any strength will suppress $T_c$ initially.   The increase in
$T_c$ with respect to the homogeneous system is slightly less than
4\%. After we increase $V$ to a significantly larger value, for
example $20t$, the $d$-wave superconductivity still survives and
the critical temperature saturates. This is consistent with the
BCS theory of pair breaking by point like impurities of a d-wave
superconductor (without correlations in the particle-hole
channel), where increasing impurity potential past the bandwidth
($\sim4t$) drives the impurity into the unitarity limit where the
scattering rate saturates\cite{p_hirschfeld_96}. Increasing the
impurity concentration beyond 3\% causes a dramatic monotonic drop
in $T_c$ for all $V > 0$. For 6\% impurity concentration $T_c$
vanishes even before $V=2t$.

\begin{figure}[ht]
\includegraphics[width=3.2in, clip = true]{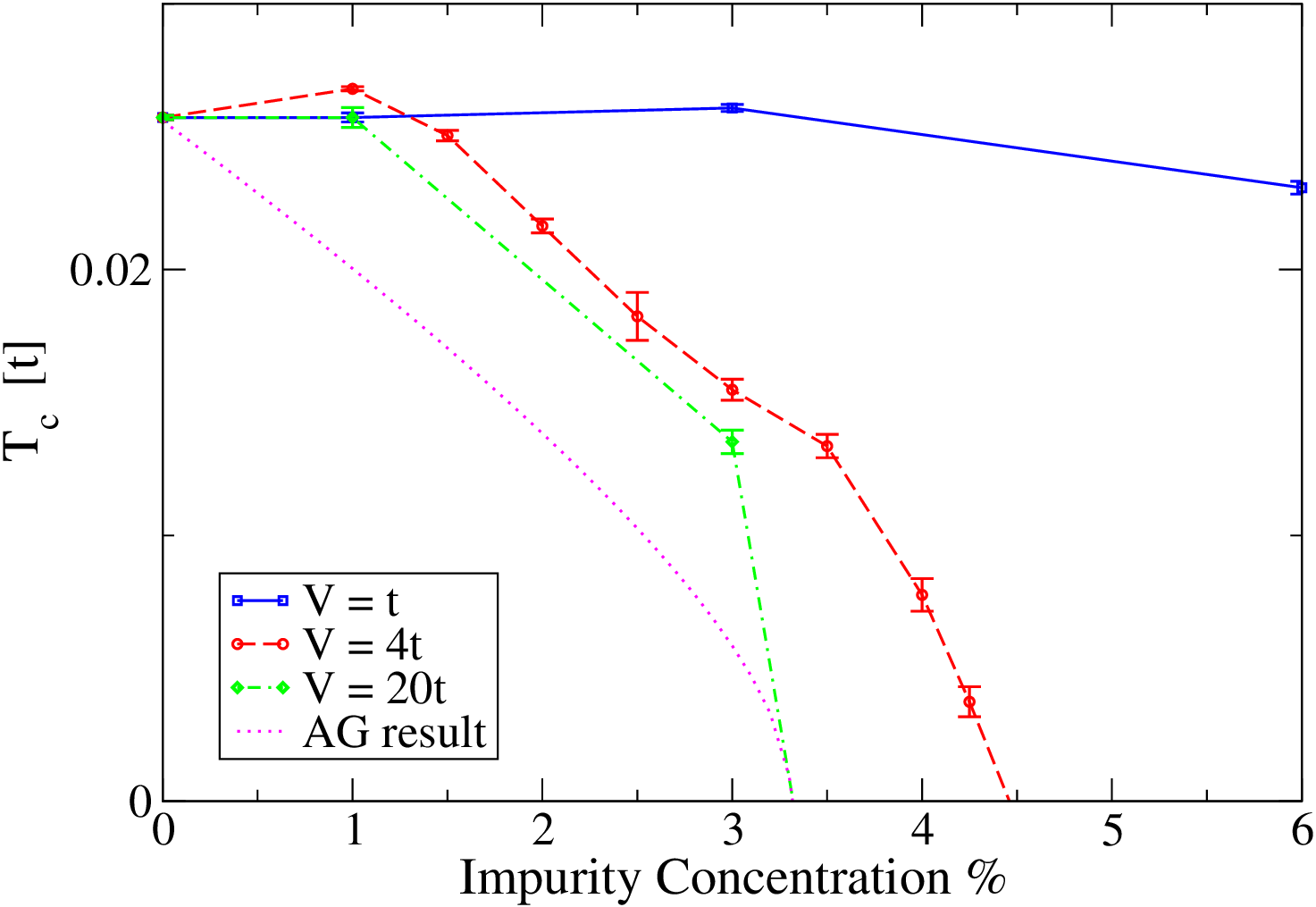}
\caption{The critical temperature $T_c$ as a function of impurity concentration for $N_c$ = 16,
and $U$ = $4t$, at impurity potentials $V = t$, $V = 4t$ and $V = 20t$. Error bars
are calculated from the extrapolation of the pair-field susceptibility\cite{t_maier_02b}.
The AG result is a fit to the critical concentration for $V = 20t$.}
\label{fig:Tcx}
\end{figure}

Fig. \ref{fig:Tcx} shows the behavior of $T_c$ as a function of
impurity concentration. For all concentrations considered, the
initial slope is either positive or nearly zero,   in marked
contrast to the  Abrikosov\textendash Gor'kov curve, which has a
negative initial slope for any combination of impurity
concentration and potential. The AG curve plotted was obtained by
fitting the unknown parameters to the critical concentration for
$V = 20t$, thus forcing the curve to go through the critical
concentration calculated by the DCA. While for a given $V$ and
impurity concentration we cannot make a direct calculation of the
pairbreaking parameter entering the noninteracting AG theory and
thus determine the critical concentration independently, the
qualitative differences of our results from the AG curve shown are
obvious, particularly for small concentrations.
%
The critical concentration calculated
for strong impurities agrees with the experimentally determined
concentration for Cu-substitution by magnetic- and nonmagnetic impurities
in LSCO\cite{g_xiao_90}.

\begin{figure}[ht]
\includegraphics[width=3.2in, clip = true]{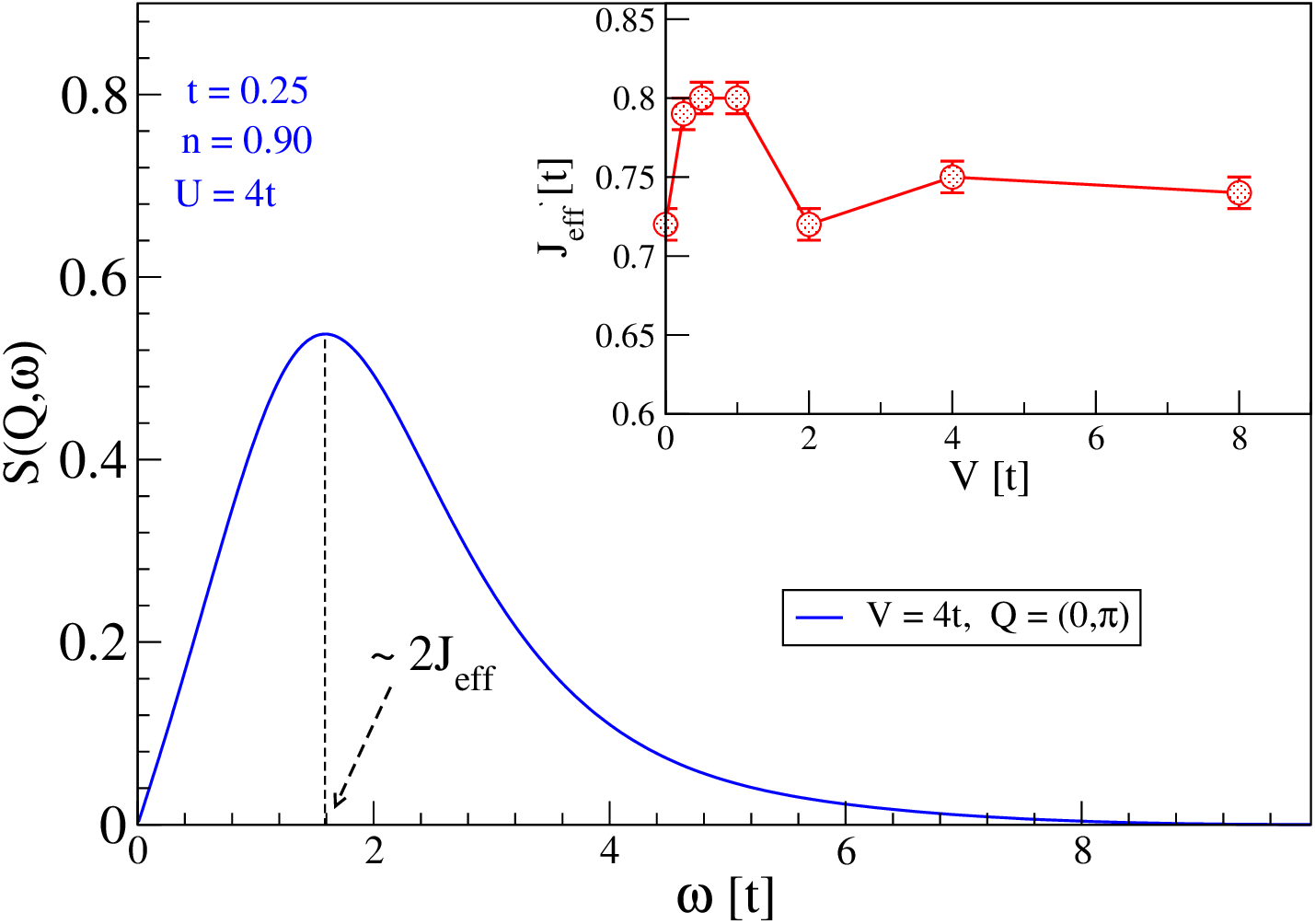}%
\caption{The dynamic spin susceptibility at $\vec{Q} = (0,\pi)$ for
$N_c$ = 16, $U$ = $4t$ and $V$ = $4t$, at impurity concentration
$x$ = 3\%.  The location of the peak, is a measure of the effective
spin-wave exchange $2\Jeff$\cite{e_manousakis_91}.
Inset: Spin coupling constant J as a function of $V$.}
\label{fig:Jv}
\end{figure}

Fig.~\ref{fig:Jv} shows the magnetic structure factor
$S(\vec{Q},\omega)$ at $\vec{Q}=(0,\pi)$ of the system at
temperature $T=0.087t$ ($\approx 3T_c$).  In analogy with linear
spin wave theory\cite{e_manousakis_91}, we note that the peak
position of $S(\vec{Q},\omega)$ at $\vec{Q}=(0,\pi)$ is a measure
of the effective exchange coupling $2\Jeff$ of the system.
Therefore, we use $S(\vec{Q}=(0,\pi),\omega)$ to extract $\Jeff$
of both ordered and disordered systems. We find that the rise of
$T_c$ at low $V$ is correlated with $\Jeff$ of the system, as
shown in the inset of Fig. \ref{fig:Jv}. The initial rise of $T_c$
tracks the initial increase in $\Jeff$. 
Then both $T_c$ and
$\Jeff$ remain nearly constant up to $V = t$. Further increase in
$V$ causes $\Jeff$ to remain roughly constant while $T_c$ shows a
dramatic drop, indicating that the suppression at higher
concentrations is indeed due to pairbreaking rather than pair
weakening.

\begin{figure}[ht]
\includegraphics[width=3.2in, clip=true]{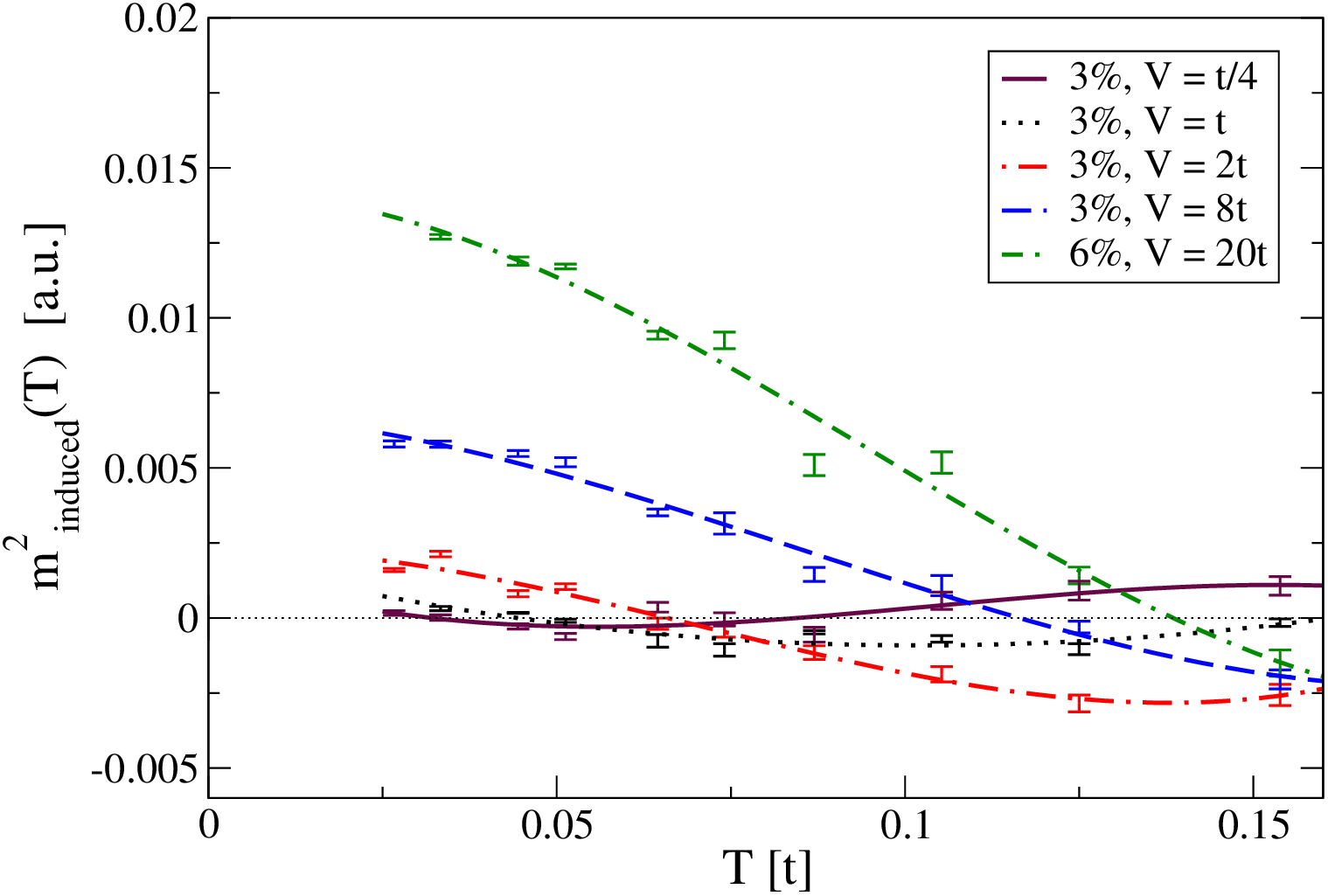}
\caption{Squared local magnetization (see Eq. \ref{eq:magnetization}) 
for $N_c$ = 16, and $U$ = $4t$, as a function of
temperature for various impurity potentials and concentrations.
The solid lines are guides to the eye.}
\label{fig:moment}
\end{figure}

For 3\% impurity concentration, $T_c$ starts to drop at $V \simeq
U/2$, which is coincident with the formation of impurity-induced
moments as shown in Fig.~\ref{fig:moment}. A weak impurity does
not induce any local moment in the system and increased $\Jeff$
causes a rise in $T_c$; increasing $V$ causes formation of local
moments and thereby enhanced pairbreaking.
%

{\em Discussion.}
Potential scattering due to weak local impurities is expected to
inhibit superconductivity, because the resulting isotropic
scattering in momentum space causes $d$-wave pair breaking and
thus a reduction in the $d$-wave order parameter. However, the
suppression found here is in general weaker than expected from
pairbreaking due to pointlike potential scatterers in a $d$-wave
system, which follows the AG form. Our results suggest that the
slowness of the initial $T_c$ suppression is due to the initial
enhancement of the interaction by the impurities.
 These results are consistent with
recent calculations\cite{m_maska_07a} where the authors argue
that an isolated
impurity in a $t-J$ model can enhance pairing locally.
Since the instantaneous part of
the pairing potential in the t-J model is proportional to
$J$\cite{t_maier_08}, the local pairing and the transition
temperature $T_c$ are enhanced. Here we have confirmed that for
impurities, $T_c$ rises along with $J$.
Intuitively, the increase in $J$ can be understood by considering
the $2^{nd}$ order exchange between two spins on sites with unequal energies,
as discussed in Ref. \onlinecite{m_maska_07a}.


We are not aware of any experimental data indicating an actual
increase or complete insensitivity of $T_c$ to increasing weak
disorder in the cuprates, when doping is held fixed. It is not
surprising, however, that our results overestimate the
 pairing enhancement effect of disorder, given the crude way in
 which disorder averaging has been performed here due to the current limitations
 on cluster size.  Nevertheless,
 we regard these results as a strong indication that the observed
 slow  initial suppression of $T_c$ in the cuprates, which has
 been remarked upon for many years, has its origin in large part
 in correlation effects.  A  point in the same general spirit was made within a
 different scheme for treating interactions in Ref.
 \onlinecite{m_kulic_97}.


{\em Conclusions.}
We have studied the effect of pointlike impurities in cuprates
using the Dynamical Cluster approximation. Our results show that
for weak local impurities, the superconducting critical
temperature $T_c$ is weakly increased due to an average,
impurity-induced enhancement of the antiferromagnetic exchange
correlation $J$. With increasing impurity strength, local moments
start to form around the impurity site, causing more quasiparticle
scattering, and the critical temperature plateaus and subsequently
decreases due to pairbreaking in both potential and magnetic
channels. The suppression of $T_c$ continues until the unitary
scattering limit is reached, and $T_c$ remains constant.

As a function of impurity concentration, $T_c$ is found to be
enhanced by or insensitive to small amounts of disorder, and
although with large disorder $T_c$ is driven to zero, the
suppression appears to be generally weaker than that predicted in
Abrikosov-Gor'kov theory, where the slope of $T_c$ versus impurity
concentration and potential is negative for all concentrations and
potentials larger than zero. Our results therefore strongly
suggest that the observed slow suppression of $T_c$ is related to
the strong correlations in the system neglected in the BCS
approach to disorder in a $d$-wave superconductor. Together with
the results of  Garg et al.\cite{a_garg_2006} and Andersen et
al\cite{b_andersen_07}, our work suggests a robustness of
superconductivity in the presence of correlations against weak
disorder in the charge channel.

\begin{acknowledgments}
This work was supported by DOE grants DE-FG02-02ER45995,
DE-FG02-97ER45660, DE-FG02-05ER46236 and NSF grant DMR-0706379. A
portion of this research at Oak Ridge National Laboratory's Center
for Nanophase Materials Sciences was sponsored by the Scientific
User Facilities Division, Office of Basic Energy Sciences, U.S.
Department of Energy. The authors acknowledge the University of
Florida High-Performance Computing Center for providing
computational
support.
\end{acknowledgments}

\bibliography{master}

\end{document}